\documentclass[12pt]{article}
\def\PsfigVersion{1.9}

\ifx\undefined\psfig\else\endinput\fi

\let\LaTeXAtSign=\@
\let\@=\relax
\edef\psfigRestoreAt{\catcode`\@=\number\catcode`@\relax}
\catcode`\@=11\relax
\newwrite\@unused
\def\ps@typeout#1{{\let\protect\string\immediate\write\@unused{#1}}}
\ps@typeout{psfig/tex \PsfigVersion}

\def\figurepath{./}

\def\@nnil{\@nil}
\def\@empty{}
\def\@psdonoop#1\@@#2#3{}
\def\@psdo#1:=#2\do#3{\edef\@psdotmp{#2}\ifx\@psdotmp\@empty \else
    \expandafter\@psdoloop#2,\@nil,\@nil\@@#1{#3}\fi}
\def\@psdoloop#1,#2,#3\@@#4#5{\def#4{#1}\ifx #4\@nnil \else
       #5\def#4{#2}\ifx #4\@nnil \else#5\@ipsdoloop #3\@@#4{#5}\fi\fi}
\def\@ipsdoloop#1,#2\@@#3#4{\def#3{#1}\ifx #3\@nnil 
       \let\@nextwhile=\@psdonoop \else
      #4\relax\let\@nextwhile=\@ipsdoloop\fi\@nextwhile#2\@@#3{#4}}
\def\@tpsdo#1:=#2\do#3{\xdef\@psdotmp{#2}\ifx\@psdotmp\@empty \else
    \@tpsdoloop#2\@nil\@nil\@@#1{#3}\fi}
\def\@tpsdoloop#1#2\@@#3#4{\def#3{#1}\ifx #3\@nnil 
       \let\@nextwhile=\@psdonoop \else
      #4\relax\let\@nextwhile=\@tpsdoloop\fi\@nextwhile#2\@@#3{#4}}
\ifx\undefined\fbox
\newdimen\fboxrule
\newdimen\fboxsep
\newdimen\ps@tempdima
\newbox\ps@tempboxa
\long\def\fbox#1{\leavevmode\setbox\ps@tempboxa\hbox{#1}\ps@tempdima\fboxrule
    \advance\ps@tempdima \fboxsep \advance\ps@tempdima \dp\ps@tempboxa
   \hbox{\lower \ps@tempdima\hbox
  {\vbox{\hrule height \fboxrule
          \hbox{\vrule width \fboxrule \hskip\fboxsep
          \vbox{\vskip\fboxsep \box\ps@tempboxa\vskip\fboxsep}\hskip 
                 \fboxsep\vrule width \fboxrule}
                 \hrule height \fboxrule}}}}
\fi
\newread\ps@stream
\newif\ifnot@eof       
\newif\if@noisy        
\newif\if@atend        
\newif\if@psfile       
{\catcode`\%=12\global\gdef\epsf@start{
\def\epsf@PS{PS}
\def\epsf@getbb#1{%
\openin\ps@stream=#1
\ifeof\ps@stream\ps@typeout{Error, File #1 not found}\else
   {\not@eoftrue \chardef\other=12
    \def\do##1{\catcode`##1=\other}\dospecials \catcode`\ =10
    \loop
       \if@psfile
	  \read\ps@stream to \epsf@fileline
       \else{
	  \obeyspaces
          \read\ps@stream to \epsf@tmp\global\let\epsf@fileline\epsf@tmp}
       \fi
       \ifeof\ps@stream\not@eoffalse\else
       \if@psfile\else
       \expandafter\epsf@test\epsf@fileline:. \\%
       \fi
          \expandafter\epsf@aux\epsf@fileline:. \\%
       \fi
   \ifnot@eof\repeat
   }\closein\ps@stream\fi}%
\long\def\epsf@test#1#2#3:#4\\{\def\epsf@testit{#1#2}
			\ifx\epsf@testit\epsf@start\else
\ps@typeout{Warning! File does not start with `\epsf@start'.  It may not be a 
PostScript file.}
			\fi
			\@psfiletrue} 
{\catcode`\%=12\global\let\epsf@percent=
\long\def\epsf@aux#1#2:#3\\{\ifx#1\epsf@percent
   \def\epsf@testit{#2}\ifx\epsf@testit\epsf@bblit
	\@atendfalse
        \epsf@atend #3 . \\%
	\if@atend	
	   \if@verbose{
		\ps@typeout{psfig: found `(atend)'; continuing search}
	   }\fi
        \else
        \epsf@grab #3 . . . \\%
        \not@eoffalse
        \global\no@bbfalse
        \fi
   \fi\fi}%
\def\epsf@grab #1 #2 #3 #4 #5\\{%
   \global\def\epsf@llx{#1}\ifx\epsf@llx\empty
      \epsf@grab #2 #3 #4 #5 .\\\else
   \global\def\epsf@lly{#2}%
   \global\def\epsf@urx{#3}\global\def\epsf@ury{#4}\fi}%
\def\epsf@atendlit{(atend)} 
\def\epsf@atend #1 #2 #3\\{%
   \def\epsf@tmp{#1}\ifx\epsf@tmp\empty
      \epsf@atend #2 #3 .\\\else
   \ifx\epsf@tmp\epsf@atendlit\@atendtrue\fi\fi}

\chardef\psletter = 11 
\chardef\other = 12

\newif \ifdebug 
\newif\ifc@mpute 
\c@mputetrue 

\let\then = \relax
\def\r@dian{pt }
\let\r@dians = \r@dian
\let\dimensionless@nit = \r@dian
\let\dimensionless@nits = \dimensionless@nit
\def\internal@nit{sp }
\let\internal@nits = \internal@nit
\newif\ifstillc@nverging
\def \Mess@ge #1{\ifdebug \then \message {#1} \fi}

{ 
	\catcode `\@ = \psletter
	\gdef \nodimen {\expandafter \n@dimen \the \dimen}
	\gdef \term #1 #2 #3%
	       {\edef \t@ {\the #1}
		\edef \t@@ {\expandafter \n@dimen \the #2\r@dian}%
		\t@rm {\t@} {\t@@} {#3}%
	       }
	\gdef \t@rm #1 #2 #3%
	       {{%
		\count 0 = 0
		\dimen 0 = 1 \dimensionless@nit
		\dimen 2 = #2\relax
		\Mess@ge {Calculating term #1 of \nodimen 2}%
		\loop
		\ifnum	\count 0 < #1
		\then	\advance \count 0 by 1
			\Mess@ge {Iteration \the \count 0 \space}%
			\Multiply \dimen 0 by {\dimen 2}%
			\Mess@ge {After multiplication, term = \nodimen 0}%
			\Divide \dimen 0 by {\count 0}%
			\Mess@ge {After division, term = \nodimen 0}%
		\repeat
		\Mess@ge {Final value for term #1 of 
				\nodimen 2 \space is \nodimen 0}%
		\xdef \Term {#3 = \nodimen 0 \r@dians}%
		\aftergroup \Term
	       }}
	\catcode `\p = \other
	\catcode `\t = \other
	\gdef \n@dimen #1pt{#1} 
}

\def \Divide #1by #2{\divide #1 by #2} 

\def \Multiply #1by #2
       {{
	\count 0 = #1\relax
	\count 2 = #2\relax
	\count 4 = 65536
	\Mess@ge {Before scaling, count 0 = \the \count 0 \space and
			count 2 = \the \count 2}%
	\ifnum	\count 0 > 32767 
	\then	\divide \count 0 by 4
		\divide \count 4 by 4
	\else	\ifnum	\count 0 < -32767
		\then	\divide \count 0 by 4
			\divide \count 4 by 4
		\else
		\fi
	\fi
	\ifnum	\count 2 > 32767 
	\then	\divide \count 2 by 4
		\divide \count 4 by 4
	\else	\ifnum	\count 2 < -32767
		\then	\divide \count 2 by 4
			\divide \count 4 by 4
		\else
		\fi
	\fi
	\multiply \count 0 by \count 2
	\divide \count 0 by \count 4
	\xdef \product {#1 = \the \count 0 \internal@nits}%
	\aftergroup \product
       }}

\def\r@duce{\ifdim\dimen0 > 90\r@dian \then   
		\multiply\dimen0 by -1
		\advance\dimen0 by 180\r@dian
		\r@duce
	    \else \ifdim\dimen0 < -90\r@dian \then  
		\advance\dimen0 by 360\r@dian
		\r@duce
		\fi
	    \fi}

\def\Sine#1%
       {{%
	\dimen 0 = #1 \r@dian
	\r@duce
	\ifdim\dimen0 = -90\r@dian \then
	   \dimen4 = -1\r@dian
	   \c@mputefalse
	\fi
	\ifdim\dimen0 = 90\r@dian \then
	   \dimen4 = 1\r@dian
	   \c@mputefalse
	\fi
	\ifdim\dimen0 = 0\r@dian \then
	   \dimen4 = 0\r@dian
	   \c@mputefalse
	\fi
	\ifc@mpute \then
		\divide\dimen0 by 180
		\dimen0=3.141592654\dimen0
		\dimen 2 = 3.1415926535897963\r@dian 
		\divide\dimen 2 by 2 
		\Mess@ge {Sin: calculating Sin of \nodimen 0}%
		\count 0 = 1 
		\dimen 2 = 1 \r@dian 
		\dimen 4 = 0 \r@dian 
		\loop
			\ifnum	\dimen 2 = 0 
			\then	\stillc@nvergingfalse 
			\else	\stillc@nvergingtrue
			\fi
			\ifstillc@nverging 
			\then	\term {\count 0} {\dimen 0} {\dimen 2}%
				\advance \count 0 by 2
				\count 2 = \count 0
				\divide \count 2 by 2
				\ifodd	\count 2 
				\then	\advance \dimen 4 by \dimen 2
				\else	\advance \dimen 4 by -\dimen 2
				\fi
		\repeat
	\fi		
			\xdef \sine {\nodimen 4}%
       }}

\def\Cosine#1{\ifx\sine\UnDefined\edef\Savesine{\relax}\else
		             \edef\Savesine{\sine}\fi
	{\dimen0=#1\r@dian\advance\dimen0 by 90\r@dian
	 \Sine{\nodimen 0}
	 \xdef\cosine{\sine}
	 \xdef\sine{\Savesine}}}	      

\def\psdraft{
	\def\@psdraft{0}
}
\def\psfull{
	\def\@psdraft{100}
}

\psfull

\newif\if@scalefirst
\def\psscalefirst{\@scalefirsttrue}
\def\psrotatefirst{\@scalefirstfalse}
\psrotatefirst

\newif\if@draftbox
\def\psnodraftbox{
	\@draftboxfalse
}
\def\psdraftbox{
	\@draftboxtrue
}
\@draftboxtrue

\newif\if@prologfile
\newif\if@postlogfile
\def\pssilent{
	\@noisyfalse
}
\def\psnoisy{
	\@noisytrue
}
\psnoisy
\newif\if@bbllx
\newif\if@bblly
\newif\if@bburx
\newif\if@bbury
\newif\if@height
\newif\if@width
\newif\if@rheight
\newif\if@rwidth
\newif\if@angle
\newif\if@clip
\newif\if@verbose
\def\@p@@sclip#1{\@cliptrue}

\newif\if@decmpr

\def\@p@@sfigure#1{\def\@p@sfile{null}\def\@p@sbbfile{null}
	        \openin1=#1.bb
		\ifeof1\closein1
	        	\openin1=\figurepath#1.bb
			\ifeof1\closein1
			        \openin1=#1
				\ifeof1\closein1%
				       \openin1=\figurepath#1
					\ifeof1
					   \ps@typeout{Error, File #1 not found}
						\if@bbllx\if@bblly
				   		\if@bburx\if@bbury
			      				\def\@p@sfile{#1}%
			      				\def\@p@sbbfile{#1}%
							\@decmprfalse
				  	   	\fi\fi\fi\fi
					\else\closein1
				    		\def\@p@sfile{\figurepath#1}%
				    		\def\@p@sbbfile{\figurepath#1}%
						\@decmprfalse
	                       		\fi%
			 	\else\closein1%
					\def\@p@sfile{#1}
					\def\@p@sbbfile{#1}
					\@decmprfalse
			 	\fi
			\else
				\def\@p@sfile{\figurepath#1}
				\def\@p@sbbfile{\figurepath#1.bb}
				\@decmprtrue
			\fi
		\else
			\def\@p@sfile{#1}
			\def\@p@sbbfile{#1.bb}
			\@decmprtrue
		\fi}

\def\@p@@sfile#1{\@p@@sfigure{#1}}

\def\@p@@sbbllx#1{
		\@bbllxtrue
		\dimen100=#1
		\edef\@p@sbbllx{\number\dimen100}
}
\def\@p@@sbblly#1{
		\@bbllytrue
		\dimen100=#1
		\edef\@p@sbblly{\number\dimen100}
}
\def\@p@@sbburx#1{
		\@bburxtrue
		\dimen100=#1
		\edef\@p@sbburx{\number\dimen100}
}
\def\@p@@sbbury#1{
		\@bburytrue
		\dimen100=#1
		\edef\@p@sbbury{\number\dimen100}
}
\def\@p@@sheight#1{
		\@heighttrue
		\dimen100=#1
   		\edef\@p@sheight{\number\dimen100}
}
\def\@p@@swidth#1{
		\@widthtrue
		\dimen100=#1
		\edef\@p@swidth{\number\dimen100}
}
\def\@p@@srheight#1{
		\@rheighttrue
		\dimen100=#1
		\edef\@p@srheight{\number\dimen100}
}
\def\@p@@srwidth#1{
		\@rwidthtrue
		\dimen100=#1
		\edef\@p@srwidth{\number\dimen100}
}
\def\@p@@sangle#1{
		\@angletrue
		\edef\@p@sangle{#1} 
}
\def\@p@@ssilent#1{ 
		\@verbosefalse
}
\def\@p@@sprolog#1{\@prologfiletrue\def\@prologfileval{#1}}
\def\@p@@spostlog#1{\@postlogfiletrue\def\@postlogfileval{#1}}
\def\@cs@name#1{\csname #1\endcsname}
\def\@setparms#1=#2,{\@cs@name{@p@@s#1}{#2}}
\def\ps@init@parms{
		\@bbllxfalse \@bbllyfalse
		\@bburxfalse \@bburyfalse
		\@heightfalse \@widthfalse
		\@rheightfalse \@rwidthfalse
		\def\@p@sbbllx{}\def\@p@sbblly{}
		\def\@p@sbburx{}\def\@p@sbbury{}
		\def\@p@sheight{}\def\@p@swidth{}
		\def\@p@srheight{}\def\@p@srwidth{}
		\def\@p@sangle{0}
		\def\@p@sfile{} \def\@p@sbbfile{}
		\def\@p@scost{10}
		\def\@sc{}
		\@prologfilefalse
		\@postlogfilefalse
		\@clipfalse
		\if@noisy
			\@verbosetrue
		\else
			\@verbosefalse
		\fi
}
\def\parse@ps@parms#1{
	 	\@psdo\@psfiga:=#1\do
		   {\expandafter\@setparms\@psfiga,}}
\newif\ifno@bb
\def\bb@missing{
	\if@verbose{
		\ps@typeout{psfig: searching \@p@sbbfile \space  for bounding 
box}
	}\fi
	\no@bbtrue
	\epsf@getbb{\@p@sbbfile}
        \ifno@bb \else \bb@cull\epsf@llx\epsf@lly\epsf@urx\epsf@ury\fi
}	
\def\bb@cull#1#2#3#4{
	\dimen100=#1 bp\edef\@p@sbbllx{\number\dimen100}
	\dimen100=#2 bp\edef\@p@sbblly{\number\dimen100}
	\dimen100=#3 bp\edef\@p@sbburx{\number\dimen100}
	\dimen100=#4 bp\edef\@p@sbbury{\number\dimen100}
	\no@bbfalse
}
\newdimen\p@intvaluex
\newdimen\p@intvaluey
\def\rotate@#1#2{{\dimen0=#1 sp\dimen1=#2 sp
		  \global\p@intvaluex=\cosine\dimen0
		  \dimen3=\sine\dimen1
		  \global\advance\p@intvaluex by -\dimen3
		  \global\p@intvaluey=\sine\dimen0
		  \dimen3=\cosine\dimen1
		  \global\advance\p@intvaluey by \dimen3
		  }}
\def\compute@bb{
		\no@bbfalse
		\if@bbllx \else \no@bbtrue \fi
		\if@bblly \else \no@bbtrue \fi
		\if@bburx \else \no@bbtrue \fi
		\if@bbury \else \no@bbtrue \fi
		\ifno@bb \bb@missing \fi
		\ifno@bb \ps@typeout{FATAL ERROR: no bb supplied or found}
			\no-bb-error
		\fi
		\count203=\@p@sbburx
		\count204=\@p@sbbury
		\advance\count203 by -\@p@sbbllx
		\advance\count204 by -\@p@sbblly
		\edef\ps@bbw{\number\count203}
		\edef\ps@bbh{\number\count204}
		\if@angle 
			\Sine{\@p@sangle}\Cosine{\@p@sangle}
	        	{\dimen100=\maxdimen\xdef\r@p@sbbllx{\number\dimen100}
					    \xdef\r@p@sbblly{\number\dimen100}
			                    \xdef\r@p@sbburx{-\number\dimen100}
					    \xdef\r@p@sbbury{-\number\dimen100}}
                        \def\minmaxtest{
			   \ifnum\number\p@intvaluex<\r@p@sbbllx
			      \xdef\r@p@sbbllx{\number\p@intvaluex}\fi
			   \ifnum\number\p@intvaluex>\r@p@sbburx
			      \xdef\r@p@sbburx{\number\p@intvaluex}\fi
			   \ifnum\number\p@intvaluey<\r@p@sbblly
			      \xdef\r@p@sbblly{\number\p@intvaluey}\fi
			   \ifnum\number\p@intvaluey>\r@p@sbbury
			      \xdef\r@p@sbbury{\number\p@intvaluey}\fi
			   }
			\rotate@{\@p@sbbllx}{\@p@sbblly}
			\minmaxtest
			\rotate@{\@p@sbbllx}{\@p@sbbury}
			\minmaxtest
			\rotate@{\@p@sbburx}{\@p@sbblly}
			\minmaxtest
			\rotate@{\@p@sbburx}{\@p@sbbury}
			\minmaxtest
			\edef\@p@sbbllx{\r@p@sbbllx}\edef\@p@sbblly{\r@p@sbblly}
			\edef\@p@sbburx{\r@p@sbburx}\edef\@p@sbbury{\r@p@sbbury}
		\fi
		\count203=\@p@sbburx
		\count204=\@p@sbbury
		\advance\count203 by -\@p@sbbllx
		\advance\count204 by -\@p@sbblly
		\edef\@bbw{\number\count203}
		\edef\@bbh{\number\count204}
}
\def\in@hundreds#1#2#3{\count240=#2 \count241=#3
		     \count100=\count240	
		     \divide\count100 by \count241
		     \count101=\count100
		     \multiply\count101 by \count241
		     \advance\count240 by -\count101
		     \multiply\count240 by 10
		     \count101=\count240	
		     \divide\count101 by \count241
		     \count102=\count101
		     \multiply\count102 by \count241
		     \advance\count240 by -\count102
		     \multiply\count240 by 10
		     \count102=\count240	
		     \divide\count102 by \count241
		     \count200=#1\count205=0
		     \count201=\count200
			\multiply\count201 by \count100
		 	\advance\count205 by \count201
		     \count201=\count200
			\divide\count201 by 10
			\multiply\count201 by \count101
			\advance\count205 by \count201
		     \count201=\count200
			\divide\count201 by 100
			\multiply\count201 by \count102
			\advance\count205 by \count201
		     \edef\@result{\number\count205}
}
\def\compute@wfromh{
		\in@hundreds{\@p@sheight}{\@bbw}{\@bbh}
		\edef\@p@swidth{\@result}
}
\def\compute@hfromw{
	        \in@hundreds{\@p@swidth}{\@bbh}{\@bbw}
		\edef\@p@sheight{\@result}
}
\def\compute@handw{
		\if@height 
			\if@width
			\else
				\compute@wfromh
			\fi
		\else 
			\if@width
				\compute@hfromw
			\else
				\edef\@p@sheight{\@bbh}
				\edef\@p@swidth{\@bbw}
			\fi
		\fi
}
\def\compute@resv{
		\if@rheight \else \edef\@p@srheight{\@p@sheight} \fi
		\if@rwidth \else \edef\@p@srwidth{\@p@swidth} \fi
}
\def\compute@sizes{
	\compute@bb
	\if@scalefirst\if@angle
	\if@width
	   \in@hundreds{\@p@swidth}{\@bbw}{\ps@bbw}
	   \edef\@p@swidth{\@result}
	\fi
	\if@height
	   \in@hundreds{\@p@sheight}{\@bbh}{\ps@bbh}
	   \edef\@p@sheight{\@result}
	\fi
	\fi\fi
	\compute@handw
	\compute@resv}

\def\psfig#1{\vbox {
	\ps@init@parms
	\parse@ps@parms{#1}
	\compute@sizes
	\ifnum\@p@scost<\@psdraft{
		\special{ps::[begin] 	\@p@swidth \space \@p@sheight \space
				\@p@sbbllx \space \@p@sbblly \space
				\@p@sbburx \space \@p@sbbury \space
				startTexFig \space }
		\if@angle
			\special {ps:: \@p@sangle \space rotate \space} 
		\fi
		\if@clip{
			\if@verbose{
				\ps@typeout{(clip)}
			}\fi
			\special{ps:: doclip \space }
		}\fi
		\if@prologfile
		    \special{ps: plotfile \@prologfileval \space } \fi
		\if@decmpr{
			\if@verbose{
				\ps@typeout{psfig: including \@p@sfile.Z \space 
}
			}\fi
			\special{ps: plotfile "`zcat \@p@sfile.Z" \space }
		}\else{
			\if@verbose{
				\ps@typeout{psfig: including \@p@sfile \space }
			}\fi
			\special{ps: plotfile \@p@sfile \space }
		}\fi
		\if@postlogfile
		    \special{ps: plotfile \@postlogfileval \space } \fi
		\special{ps::[end] endTexFig \space }
		\vbox to \@p@srheight sp{
			\hbox to \@p@srwidth sp{
				\hss
			}
		\vss
		}
	}\else{
		\if@draftbox{		
			\hbox{\frame{\vbox to \@p@srheight sp{
			\vss
			\hbox to \@p@srwidth sp{ \hss \@p@sfile \hss }
			\vss
			}}}
		}\else{
			\vbox to \@p@srheight sp{Return-Path: Karl.Joulain 
Return-Path: <Karl.Joulain>
Received: from localhost (localhost [127.0.0.1])
          by lra.ens.fr (8.8.4/8.8.4) with SMTP
	  id MAA14401 for thoraval; Wed, 11 Mar 1998 12:34:46 +0100 (MET)
From: Karl Joulain <Karl.Joulain>
Message-Id: <199803111134.MAA14401@lra.ens.fr>
X-Authentication-Warning: lra.ens.fr: localhost [127.0.0.1] didn't use HELO protocol
To: thoraval
Date: Wed, 11 Mar 98 12:34:46 +0100
X-Mts: smtp

			\vss
			\hbox to \@p@srwidth sp{\hss}
			\vss
			}
		}\fi

	}\fi
}}
\psfigRestoreAt
\let\@=\LaTeXAtSign

\usepackage{amssymb}
\usepackage{latexsym}
\newcommand{\bea}{\begin{eqnarray}}
\newcommand{\be}{\begin{equation}}
\newcommand{\eea}{\end{eqnarray}}
\newcommand{\ee}{\end{equation}}
\newcommand{\df}[2]{\displaystyle\frac {#1}{#2}}
\def\nn{\nonumber}
\def\le{\left}
\def\ri{\right}
\def\disp{\displaystyle}
\def\part{\partial}
\def\td{\tilde}
\def\w{\wedge}
\def\a{\alpha}
\def\b{\beta}
\def\d{\delta}
\def\e{\epsilon}
\def\f{\phi}
\def\g{\gamma}
\def\k{\kappa}
\def\io{\iota}
\def\ps{\psi}
\def\l{\lambda}
\def\m{\mu} 
\def\n{\nu}
\def\o{\omega}
\def\p{\pi}
\def\th{\theta}
\def\r{\rho}
\def\s{\sigma}
\def\t{\tau}
\def\u{\upsilon}
\def\x{\xi}
\def\z{\zeta}
\def\D{\Delta}
\def\F{\Phi}
\def\G{\Gamma}
\def\J{\Psi}
\def\L{\Lambda}
\def\O{\Omega}
\def\P{\Pi}
\def\Q{\Theta}
\def\U{\Upsilon}
\def\X{\Xi}
\def\Si{\Sigma}
\def\T{\Theta}
\begin{document}
\title{On the time evolution of the redshift and the luminosity distance}
\author{C. Barbachoux\thanks{e-mail: barba@ccr.jussieu.fr}, G. Le Denmat\thanks{e-mail: gele@ccr.jussieu.fr} \\
{\small LERMA/UMR 8540, Universit\'e Pierre et Marie Curie,
ERGA,}\\
{\small Bo\^{\i}te 142, 4 place Jussieu, 75005 Paris Cedex 05, France.}}
\maketitle
\begin{abstract}
We investigate the temporal evolution of the redshift and the luminosity distance within the standard Friedmann-Roberston-Walker cosmological model. The redshift and luminosity distance of sources evolve with time and we show that they tend to given values, namely the stable equilibrium states associated to the first order differential equation they verify. This suggests that the sources concentrate at about these values. Furthermore, as these values depend only on the cosmological parameters, their measure could provide a new approach to determine the value of the cosmological constant. 
\end{abstract}
\section{Introduction}
Much attention has been devoted to 
the study of the temporal evolution of the redshift and the luminosity distance, because of the alternative scheme they constitute to determine the values of the cosmological parameters in comparison with other approaches  (White, Scoot \& Silk 1994; Perlmutter et al 1998; Schmidt et al 1998; Riess et al 1998). First proposed by Sandage (Sandage 1962) in a pioneering work (including an important Appendix from MacVittie), this possibility has been recently subject to a renewal of interest with the improvement of
 spectroscopic techniques (Loeb 1998). 
In this paper, our aim is to highlight some points on the temporal evolution of the redshift and the luminosity distance which have not been submitted yet to full scruting in the litterature. Observing a source at different instants, we establish that its redshift and luminosity distance evolute with time and tend to given values, namely the stable equilibrium states associated to the first order differential equation they verify. More precisely, this result brings out the existence of a specific value of the redshift to which the  sources concentrate. This can shed some light to the observation of a peak in the count of quasars at $z\sim 2.5$ (Fan et al 1999; and references therein), which up to now failed to be explained completely (Peebles \& Ratra 2002). Besides, these values depending only on the cosmological parameters, their experimental measurements could provide a new tool to determine the value of the cosmological parameter.  

\section{Time derivatives of the redshift and the luminosity distance}
Let us introduce some fundamentals in order to precise our notations.
Taking the convention $G=c=1$, we start with the Robertson-Walker line element under the form:
\be
ds^2=dt^2-a^2(t)\le(\frac {dr^2}{1-k r^2}+r^2 d\O^2\ri),
\label{e1}
\ee
with $k$ the curvature index taking the discrete values $-1$, $0$ or $+1$ and $d\O^2=d\th^2+\,sin^2\th \,d\phi^2$ the line element of the unit two-sphere.  
The cosmological redshift $z$ is given by:
\be
z+1=\frac {a_o}{a_e},
\label{e2}
\ee
with $a_o$ and $a_e$ the values of $a(t)$ at the instants $t_o$ and $t_e$, when the signal is, respectively, observed and emitted. 
In the same vein, the luminosity distance $d_L$ reads (e.g. Weinberg 1972; Peacock 2000):
\be
d_{L}=r_e a_o(1+z),
\label{e3}
\ee
with $r_e$ the radial coordinate of the source.
The apparent luminosity $l$ can be related to the luminosity distance by:
\be
l=\frac {L}{4\pi} \frac {1}{d_L^2},
\label{e4}
\ee
where $L$ is the global luminosity which is a constant. Finally, the bolometric magnitude $m$ is defined by:
\be 
m=5 \,log_{10}\,d_{L}+C
\label{e5}
\ee
where $C$ is a constant depending on the absolute magnitude and on the choice of the units of $d_L$.

Reminding that $a(t)({\part}/{\part t})$ is a conformal Killing vector field  for the metric (\ref{e1}), we obtain the relation:
\be
\frac {dt_e}{a_e}=\frac {dt_o}{a_o},
\label{t0}
\ee
which relates the intervals of proper time along the world lines of the observer and the  source.
From Equations (\ref{e2}) and (\ref{t0}), the derivative of the redshift  with respect to the time of observation is:
\be
\dot z=H_o(1+z)\le(1-h(z)\ri),
\label{t1}
\ee
where $h(z)=\df{\psi}{1+z}$ with $\psi=\df{a_o}{\dot a_o}\df{\dot a_e}{a_e}$
and $H_0$ represents the Hubble constant.
From now on, we assume a comoving source, i.e. we suppose that the effects of the peculiar velocity and acceleration (Phillipps 1982) can be neglected or at least isolated (Loeb 1998).

Using (\ref{t1}), the time derivative of the luminosity distance (\ref{e3}) reads:
\be
\dot d_{L}=2 r_e a_o H_o(1+z)\le(1-\df{1}{2}h(z)\ri).
\label{t2}
\ee
According to (\ref{e4}) and (\ref{e5}), the time derivatives of the apparent luminosity and the bolometric magnitude are related to the time derivative of the luminosity distance (\ref{t2}) by:
\bea
\dot l&=&- \le({\frac {16\pi l^3}{L}}\ri)^{1/2}\dot d_{L},
\label{t3}\\
\dot m&=&5\,\frac {\dot d_{L}}{d_{L}}.
\label{t4}
\eea

Using Einstein's equations, the function $\psi$ can be related to the cosmological parameters. Considering the matter content of the universe to be a perfect fluid with a mass density $\rho$ and a pressure $p$, the Einstein equations for the metric (\ref{e1}) read:
\bea
\le(\frac {\dot a}{a}\ri)^2&=&-\frac {k}{a^2}+\frac {\L}{3}+\frac {8\pi}{3}\rho,
\label{t5}\\
2\,\frac {\ddot a}{a}&=&-\le(\frac {\dot a}{a}\ri)^2-\frac {k}{a^2}+\L-8\pi p,
\label{t6}
\eea
with $\L$ the cosmological constant. 
Assuming that the fluid is a mixture of noninteracting dust and radiation, the energy momentum conservation leads to:
\bea
\rho&=&  \rho_{m\,o}\le(\frac {a_o}{a}\ri)^3+\rho_{r\,o}\le(\frac {a_o}{a}\ri)^4,
\label{t7}\\
p&=& \frac {1}{3}\rho_{r\,o}\le(\frac {a_o}{a}\ri)^4,
\label{t8}
\eea
with $\rho_m$ the matter density and $\rho_r$ the radiation density.
The combination of (\ref{t5}) and (\ref{t6}) with the aid of (\ref{t7}) and (\ref{t8}) furnishes the following relations between the cosmological parameters:
\bea
\frac{k}{a_o^2}&=&H_o^2\le(-1+\frac {3}{2}\O_{m\,o}-q_o+2\O_{r\,o}\ri),
\label{t9}\\
q_o&=&\O_{m\,o}+\frac {1}{2}\O_{r\,o}-\O_{\L\,o},
\label{t10}
\eea
where the standard notations for the deceleration parameter $q_o=-\df{\ddot a_o}{a_o}\df{1}{H_o^2}$, the matter density parameter $\O_{m\,o}=\df{8\pi}{3}\df{\rho_{m\,o}}{H_o^2}$, the radiation density parameter $\O_{r\,o}=\df{8\pi}{3}\df{\rho_{r\,o}}{H_o^2}$ and the cosmological parameter $\O_{\L\,o}=\df{1}{3}\df{\L}{H_o^2}$ have been introduced. 

Substituting (\ref{t7}) into (\ref{t5})
and appealing to the relations (\ref{t7}) and (\ref{t8}) yield for $h$ (see Phillipps 1982; Lake 1981; R\"udiger 1980; Janis, 1992) and references therein):
\be
(h(z))^2=\frac {\psi^2}{(1+z)^2}=1-\O_{\L\,o}+\O_{m\,o} z+\frac {\O_{\L\,o}}{(1+z)^2}+\O_{r\,o}z(2+z).
\label{t11}
\ee

\section{Stability of the equilibrium states of $\dot z$ and $\dot d_L$}
In the following, the observations being performed within the matter dominated aera, we assume that the radiation energy density is negligible, i.e. $\O_{r\,o}\sim 0$. Furthermore, in order to preserve generality and to grasp better the role of the cosmological constant on the evolution of the redshift and the distance luminosity, we do not impose any specific relation on the values of $\O_{m\,o}$ and $\O_{\L\,o}$. The typical cases  $(\O_{m\,o}=0.3,\O_{\L\,o}=0)$ and $(\O_{m\,o}=0.3,\O_{\L\,o}=0.7)$ are presented as illustrative examples (see figures 1-2). 

Introducing
\be
v(z)=H_o(1+z)f(z),
\label{r1}
\ee
with $f(z)=1-h(z)$ and $h(z)$ given by (\ref{t11}), Equation  (\ref{t1}) becomes 
\be
\dot z=v(z).
\label{r1b}
\ee
The equilibrium states associated to (\ref{r1b}), i.e. the values of $z\geq 0$ such that $v(z)=0$, are given by the roots of $f$:
\bea
z_{eq1}&=&0,\label{r2}\\
z_{eq2}&=&-1+\df{1}{2}\le(X+\sqrt{X(X+4)}\ri),\label{r3}\\
z_{eq3}&=&-1+\df{1}{2}\le(X-\sqrt{X(X+4)}\ri),\label{r4}
\eea
where $X=\df{\O_{\L\,o}}{\O_{m\,o}}$. The root $z_{eq2}$ is only positive for  $X\geq \df{1}{2}$ and $z_{eq3}$ is always negative. Therefore, for $X\leq \df{1}{2}$, the only equilibrium state is $z_{eq1}=0$ and for $X\geq \df{1}{2}$, the equilibrium states $z_{eq1}$ and $z_{eq2}$ have to be considered. 

The stability of these states stems from the study of the sign of $v$ (e.g. Arnold 1978).  According to (\ref{r1}), the sign of $v$ is derived from the sign of $f$. Using (\ref{t11}), the derivative of $f$ with respect to $z$ reads:
\be
\df{df}{dz}=-\df{\O_{mo}}{h(z)}\le[\df{1}{2}-\df{X}{(1+z)^3}\ri].
\label{r5}
\ee
For $X\leq \df{1}{2}$, Equation (\ref{r5}) provides $\df{df}{dz}\leq 0$ for all $z\geq 0$. The function $f$ is then  decreasing and  as $f(0)=0$, $f(z)$ is always negative. With (\ref{r1}), this implies that $v$ is negative for all $z$. The equilibrium state $z_{eq1}=0$ is then  stable for $X\leq \df{1}{2}$. An illustrative representation of the functions $f$ and $v$ is given in fig.1 for $\O_{mo}=0.3$ and $\O_{\L o}=0$. 

When $X\geq \df{1}{2}$, we deduce from (\ref{r5}) that $\df{df}{dz}\geq 0$ for  $0\leq z\leq z_1$, with $z_1=1+(2X)^{1/3}$, and $\df{df}{dz}\leq 0$ for $z\geq z_1$. The function $f$ is then increasing for $0\leq z\leq z_1$ and decreasing for $z\geq z_1$. As $f(0)=0$, $f$ is positive for $0\leq z\leq z_{eq2}$, with $z_{eq2}$ given by (\ref{r3}), reaches a maximum at $z_1\leq z_{eq2}$ and is negative for $z\geq z_{eq2}$.
From (\ref{r1}), it follows that $v$ is positive for $0\leq z\leq z_{eq2}$ and negative for $z\geq z_{eq2}$. Hence,  for $X\geq \df{1}{2}$, $z_{eq1}=0$ is a unstable equilibrium state and $z_{eq2}$ is stable.
The functions $f$ and $v$ are displayed in fig.1 when $\O_{m\,o}=0.3$ and $\O_{\L\,o}=0.7$, i.e. for $X=7/3$.

Using (\ref{e3}) and (\ref{t2}), the time derivative of the luminosity distance becomes:
\be
\dot d_{L}=2H_od_L\tilde g(d_L),
\label{r6}
\ee
where $\tilde g(d_L)=1-\df{1}{2}\tilde h(d_L)$. The function $\tilde h(d_L)$ is deduced from the expression (\ref{t11}) of $h(z)$ by performing the
substitution of $z$ by $d_L$ in accordance with (\ref{e3}). Denoting
\be
w(d_L)=2H_od_L\tilde g(d_L),
\label{r8}
\ee
Eq (\ref{r6}) becomes $\dot d_{L}=w(d_L)$.
As $d_L\geq r_ea_o$ for $z\geq 0$, the equilibrium states associated to $w$ are defined by the roots of $\tilde g(d_L)$. The solution of $\tilde g(d_L)=0$ is:
\be
d_L^{eq}=\df{r_ea_o}{3}\le[1+u+2(1+u)\,cos\,\le(\frac {1}{3}\,tan^{-1}\sqrt{\frac {4(1+u)^6}{[2(1+u)^3-27X]^2}-1}\ri)\ri],
\label{r9}
\ee
with $u=X+\df{3}{\O_{mo}}$. The stability of this equilibrium state is deduced from the study of the sign of $w$. As $d_L\geq r_e a_o$ and with (\ref{r8}), the sign of $w$ is derived from the sign of $\tilde g(d_L)$. Using (\ref{e3}), the derivative of $\tilde g$ with respect to $d_L$ reads:
\be
\df{d\tilde g}{dd_L}=\df{1}{2r_ea_o}\df{df}{dz}.
\label{r10}
\ee
Therefore, the variations of $\tilde g$ mimic those of $f$.
When $X\leq \df{1}{2}$, $f$ is decreasing, then $\tilde g$ decreases and as $\tilde g(r_ea_o)=\df{1}{2}$, $\tilde g$ is positive for $r_ea_o\leq d_L\leq d_L^{eq}$ and negative for $d_L\geq d_L^{eq}$.
For $X\geq \df{1}{2}$, $f$ is increasing for $0\leq z\leq z_1$ ($z_1=-1+(2X)^{1/3}$) and decreasing for $z\geq z_1$. The function $\tilde g$ is then increasing for $r_ea_o\leq d_L\leq r_e a_o (1+z_1)$ and decreasing for $z\geq r_e a_o(1+z_1)$.
As $\tilde g(r_e a_o)=\df{1}{2}$, $\tilde g$ is positive for $r_e a_o\leq d_L\leq d_L^{eq}$ and negative for $d_L\geq d_L^{eq}$. 

Using (\ref{r8}), we obtain finally that $w$ is positive for $r_e a_o\leq d_L\leq d_L^{eq}$ and negative for $d_L\geq d_L^{eq}$ in both cases $X\geq \df{1}{2}$ and $X\leq \df{1}{2}$. The equilibrium state $d_L^{eq}$ is then stable.
The behavior of $w$ with respect to $z$ is displayed in fig.2 for $(\O_{m\,o}=0.3,\O_{\L\,o}=0)$ and for $(\O_{m\,o}=0.3,\O_{\L\,o}=0.7)$.

\section{Discussion}
The temporal evolution of the redshift is tied to the stability of its equilibrium states. For $X\leq \df{1}{2}$, the state $z_{eq1}=0$ is stable and the redshift tends to $z_{eq1}=0$. In other words, this suggests that the sources concentrate to form a peak at about $z_{eq1}\sim 0$. This phenomenon being not corroborated by observations, the case $X\leq \df{1}{2}$ is discarded.
For $X\geq \df{1}{2}$, among the two equilibrium states $z_{eq1}=0$ and $z_{eq2}$ defined by (\ref{r3}), only the second one is stable. Therefore, the value of the redshift evolves with time until reaching $z_{eq2}$. Taking for the values of the cosmological parameters $\O_{mo}\sim 0.3$ and $\O_{\L o}\sim 0.7$ (Perlmutter et al 1998; Schmidt et al 1998, Riess et al 1998; and references therein), we find $z_{eq2}\sim 2.1$. This result can dovetail and shed some light to the observation of a peak in the count of quasars centered at about $z\sim 2.5$ (see Fan 1999 and references therein)\footnote{The preceding observations suggested a peak at about $z\sim 2$ (Shkolovsky 1967; Petrosian, Salpeter \& Szekers 1967; Kardashev 1967; Burbidge \& Burbidge 1967).}. The small difference between the prediction and the observation can be ascribed to the experimental uncertainties. The values of the cosmological parameters we used to compute $z_{eq2}$ can eventually constitute another explanation for this difference. Furthermore, the value $z_{eq2}$ may be attained asymptotically .

These considerations suggest that the value of $X=\df{\O_{\L o}}{\O_{mo}}$ ought to be greater than $ \df{1}{2}$. It constitutes an evidence for a nonvanishing cosmological constant (see Peebles \& Ratra 2002; and references therein). Assuming $\O_{mo}\sim 0.3$, the minimal value admissible for $\O_{\L o}$ is about $0.15$ and for a flat spacetime such that $\O_{mo}+\O_{\L o}=1$, the lower bound for $\O_{\L o}$ becomes $\df{1}{3}$. Moreover, from the experimental determination of $z_{eq2}$, a value of $X=\df{\O_{\L o}}{\O_{mo}}$ could be estimated from (\ref{r3}) providing therefore a new method to determine   
the value of $\O_{\L o}$.

Let us end up with some considerations on the luminosity distance. As $d_{L}^{eq}$ defined in (\ref{r9}) is a stable equilibrium state, the luminosity distance evolves with time and tends to $d_L^{eq}$. The same conclusion can be displayed for the apparent luminosity and the bolometric magnitude. Using (\ref{t3}) and (\ref{e4}), it turns out that the apparent luminosity has a stable equilibrium state $l_e$ related to $d_L^{eq}$ by:
\be
l_e=\df{L}{4\pi}\df{1}{(d_L^{eq})^2}.
\label{r11}
\ee
With (\ref{t4}) and (\ref{e5}), the same result is proven for the bolometric magnitude with a stable equilibrium state $m_e$ of the form:
\be
m_e=5\,log_{10}\,d_{L}^{eq}+C.
\label{r12}
\ee
Therefore, the apparent luminosity and bolometric magnitude evolves in time until reaching $l_e$ and $m_e$ respectively.

The approximation of $d_L^{eq}$ provides an estimation of the redshift where a peak is expected. Equation (\ref{r9}) can be rewritten:
\be
d_L^{eq}=\df{r_e a_o}{3}\le[1+u+2(1+u)\,cos\,\phi\ri],
\label{r13}
\ee
with $u=(\O_{\L o}+3)/\O_{m\,o}$ and $\phi=\df{1}{3}\,tan^{-1}\,\le[\sqrt{\frac {4(1+u)^6}{(2(1+u)^3-27X)^2}-1}\ri]$. 
The study of the variation of $\phi$ with respect to $\O_{\L o}$ yields: 
\be
0\leq \phi\leq \phi_0,
\label{r14}
\ee
where $\phi_0$ takes the value $\phi_0\sim 2\; 10^{-2}$ for $\O_{m\,o}=0.3$. It implies that $d_L^{eq}$ can be approximated by: 
\be
d_L^{eq}\simeq r_ea_o(1+u),
\label{r15}
\ee
up to an error of about $10^{-2}r_ea_o$. For $\O_{\L o}\sim 0.7$ and $\O_{mo}\sim 0.3$, we expect the luminosity distance and bolometric magnitude to evaluate in time and to form a peak at a redshift about $u\sim 12$. Albeit these results seem to be similar to that obtained for the redshift, the lack of a class of sources with the same absolute luminosity for redshifts greater than $\sim 3$ and
the actual performances of observation (Fan 2001) may not bring out the existence of such peaks. 
\begin{center}
References
\end{center}
Arnold,V. 1978, {\it Ordinary differential equations}, MIT Press\\
Burbidge,G.R. Burbidge,E.M. 1967, ApJ 148, L107\\
Fan, X. et al. 1999, AJ 118, 1\\
Fan,X. et al. 2001, AJ 122, 2833\\
Janis,A.I. 1992, ApJ 385, L391\\
Kardashev,N. 1967, ApJ 150, L135\\
Lake,K. 1981, ApJ 247, L17\\
Loeb,A. 1998, astro-ph/ 9802122\\
Peacock,J.A. 2000, {\it Cosmological Physics}, Cambridge University Press\\
Peebles,P.J.E. Ratra,B. 2002, astro-ph/0207347\\
Perlmutter, S. et al. 1989, LBL-42230, (astro-ph/9812473)\\
Petrosian,V. Salpeter,E. Szekers,P.,1967, ApJ 147, 1222\\
Phillipps,S. 1982, ApJ 22, L123\\
Riess,A.G. et al. 1998, AJ 116, 1009\\
R\"udiger,R. 1980, ApJ 240, L384\\
Sandage,A. 1962, ApJ 136, 319\\
Schmidt,B.  et al. 1998, ApJ 507, 46\\
Shkolovsky,J. 1967, ApJ 150, L1\\
Weinberg,S. 1972 {\it Gravitation and Cosmology, principles and applications of the general theory of relativity}, John Wiley \& Sons Inc.\\
White,M. Scott,D. Silk,J. 1994, Annu.Rev.Astrophys. 32, 319--370\\
\newpage
\begin{figure}[ht]
\centerline{
\hspace{-1cm}
\psfig{file=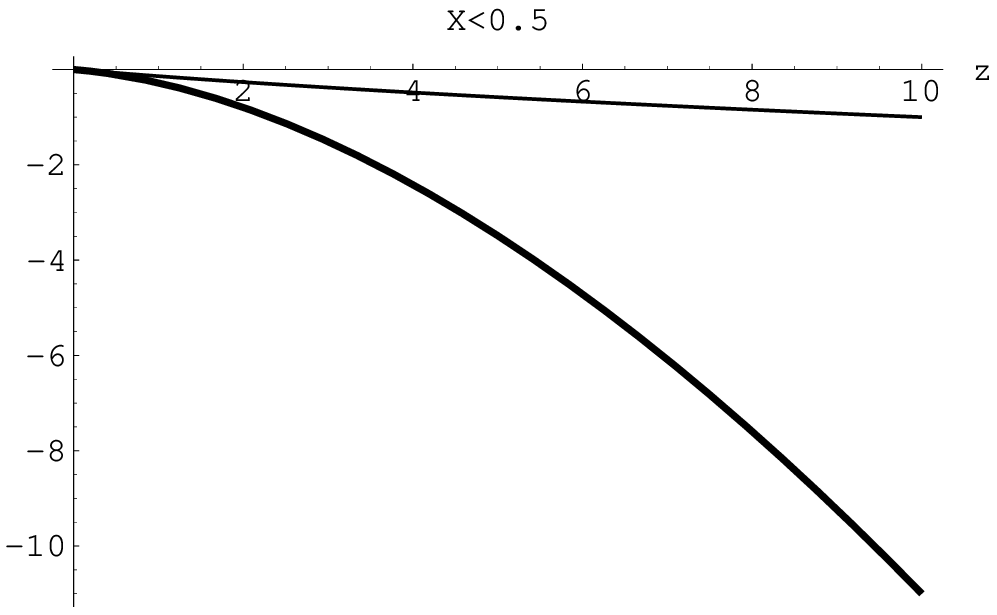,width=7cm}
\psfig{file=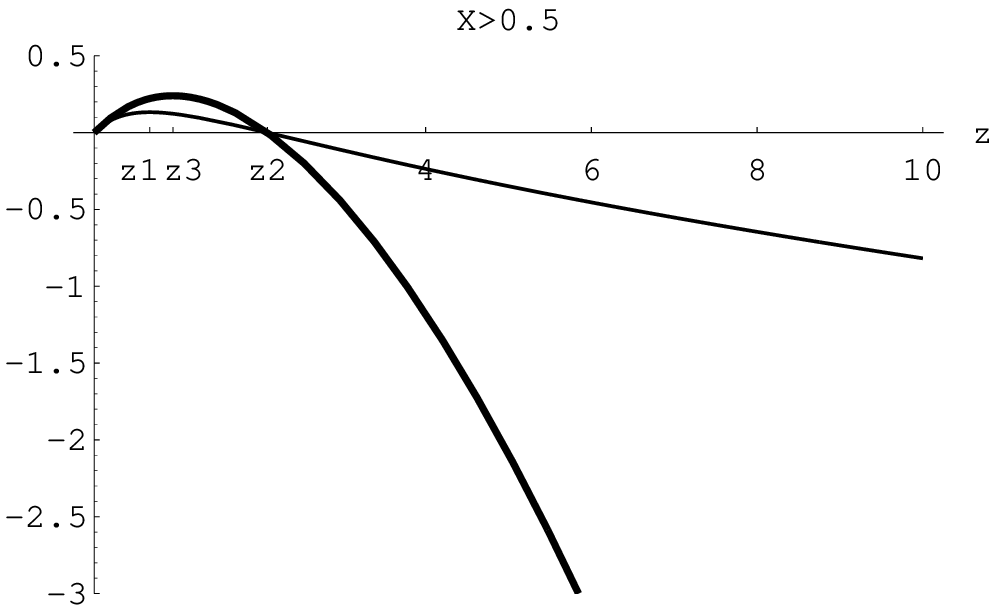,width=7cm}}
\caption{Plots of $v(z)/H_o$ (bold-faced) and $f(z)$ for $(\O_{m\,o}=0.3,\O_{\L\,o}=0)$ (i.e. $X\leq 0.5$, on the left) and for $(\O_{m\,o}=0.3,\O_{\L\,o}=0.7)$ (i.e. $X>0.5$, on the right) with $z_1=0.67$, $z_2=z_{eq2}=2.09$ and $z_3=0.95$.}
\end{figure}

\begin{figure}[ht]
\centerline{
\hspace{-1cm}
\psfig{file=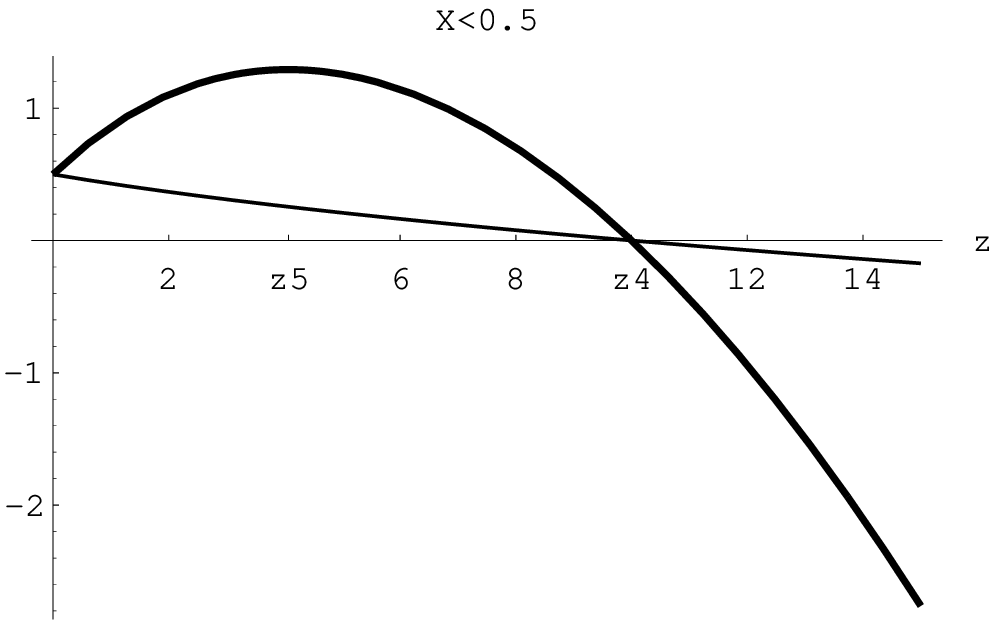,width=7cm} 
\psfig{file=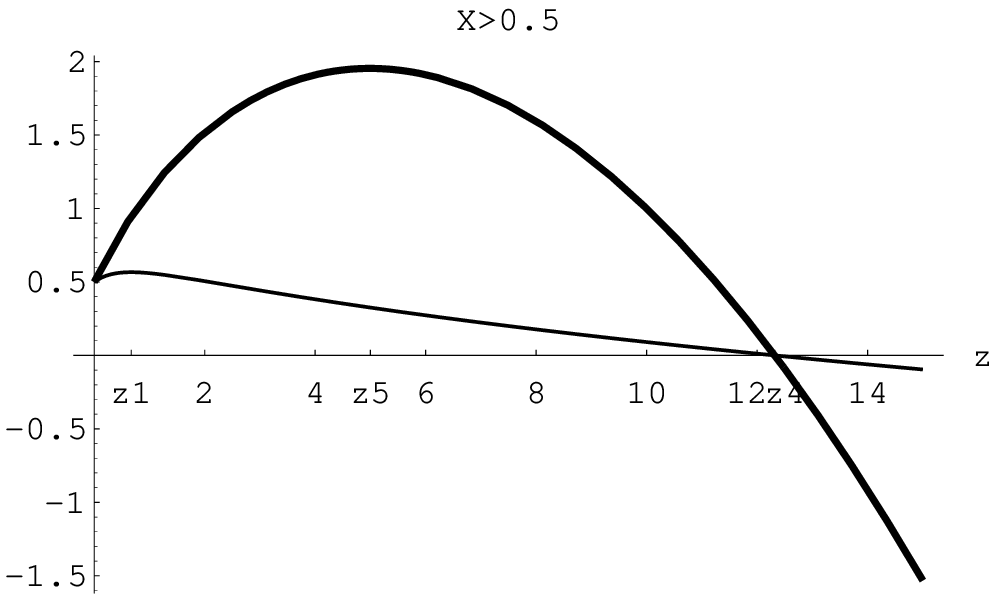,width=7cm}}
\caption{Plots of $w(z)/H_or_ea_0$ (bold-faced) and $g(z)$ for $(\O_{m\,o}=0.3,\O_{\L\,o}=0)$ (i.e. $X\leq 0.5$) with $z_4= 10$ and $z_5=4.01$ (on the left) and for $(\O_{m\,o}=0.3,\O_{\L\,o}=0.7)$ (i.e. $X>0.5$) with $z_1=0.67$, $u=12.32$ and $z_5=4.99$ (on the right).}
\end{figure}

\end{document}